# Towards a better understanding of the symmetry energy within neutron stars

C.Y. Tsang (曾浚源), M.B. Tsang (曾敏兒)[#], P. Danielewicz, and W.G. Lynch (連致標)
*National Superconducting Cyclotron Laboratory and the Department of Physics and Astronomy*
*Michigan State University, East Lansing, MI 48824 USA*
and
F.J. Fattoyev
*Center for Exploration of Energy and Matter*
*Department of Physics, Indiana University, Bloomington, IN 47408 USA*

Abstract

The LIGO-Virgo collaboration's ground-breaking detection of the binary neutron-star merger event, GW170817, has expanded efforts to understand the Equation of State (EoS) of nuclear matter. These measurements provide new constraints on the overall pressure, but do not, by itself, elucidate its microscopic origins, including the pressure arising from the symmetry energy, that governs much of the internal structure of a neutron star. To correlate microscopic constraints from nuclear measurements to the GW170817 constraints, we calculate neutron star properties with more than 200 Skyrme energy density functionals that describe properties of nuclei. Calcuated neutron-star radii *(R)* and the tidal deformabilities *(Λ)* show a strong correlation with pressure at twice saturation density. By combining the neutron star EoS extracted from the GW170817 event and the EoS of symmetric matter from nucleus-nucleus collision experiments, we extract the density dependence of the symmetry pressure from $1.2\rho_0$ to $4.5\rho_0$. While the uncertainties in the symmetry pressure are large, they can be reduced with new experimental and astrophysical results.

Email addresses:

[#]Corresponding Author: tsang@nscl.msu.edu
C.Y. Tsang: tsangc@nscl.msu.edu
W.G. Lynch: lynch@nscl.msu.edu
P. Danielewicz: danielewicz@nscl.msu.edu
F.J. Fattoyev: ffattoyev01@manhattan.edu

The Equation of State (EoS) of nuclear matter relates temperature, pressure and density of a nuclear system. It governs not only properties of nuclei and neutron stars but also the dynamics of nucleus-nucleus collisions and that of neutron-star mergers. The amount of ejected matter from the merger, which subsequently undergoes nucleosynthesis to form heavy elements up to Uranium and beyond [1-3] depends on the EoS. So does the fate of the neutron-star merger including; whether the colliding neutron stars collapse promptly into a black hole, remain a single neutron star, or form a transient neutron star that collapses later into a black hole [4]. The recent observation of a neutron-star merger event, GW170817 (GW), provides insight into the properties of nuclear matter and its equation of state (EoS) [5-9]. In this letter, we use the neutron star model in ref. [10] to explore the connection between nuclear physics experiments and neutron star properties by utilizing Skyrme interactions that are widely used to describe nuclear properties. We then explore which laboratory observables constrain the EoS at densities relevant to the neutron star tidal deformability (and radius). Finally, we focus on how the symmetry pressure at supra-saturation densities can be extracted by combining the GW astrophysical and nuclear physics experiment constraints.

In the past two decades, the nuclear EoS has been studied over a range of densities $0.25\rho_0 < \rho < 4.5\rho_0$ in nuclear structure and reaction experiments [11-23] and described with various success using ab initio [24, 25], microscopic [26, 27] and phenomenological [28-31] models. This density range is comparable to that found inside neutron stars. However the EoS from nuclear experiments using nuclei with similar number of neutrons and protons must be extrapolated to neutron star environments where the density of neutrons greatly exceeds the density of protons. Within the parabolic approximation [32], the EoS of cold nuclear matter, expressed as the energy per nucleon of the hadronic system, $\varepsilon(\rho, \delta)$, can be divided into a symmetric matter contribution, $\varepsilon(\rho, \delta=0)$, that is independent of the neutron-proton asymmetry, and a symmetry energy term, $E_{sym}(\rho,\delta)=S(\rho)\delta^2$, proportional to the square of the asymmetry, $\delta=(\rho_n-\rho_p)/\rho$, [33] as follows:

$$\varepsilon(\rho, \delta) = \varepsilon(\rho, \delta=0) + S(\rho)\delta^2 + O\delta^4 + ... \quad (1)$$

Here, $\rho_n$, $\rho_p$ and $\rho=\rho_n+\rho_p$ are the neutron, proton and nucleon densities, respectively and $S(\rho)$ is the density dependence of the symmetry energy. Relative to $S(\rho)$, the contributions to the EoS from known higher order terms, $O\delta^4$ and above, are small for $\rho<\rho_0$, less than *15%* at $2\rho_0$ and increase in importance with density [32]. Compared to current uncertainties in the EoS on neutron matter, symmetric matter, and the symmetry energy, however, these higher order terms are negligible. They will become more relevant when the uncertainties in the observational and experimental equations of state are reduced.

The extrapolation of the EoS to neutron star environments adds $P_{sym}=\rho^2 dE_{sym}(\rho,\delta)/d\rho$ to the pressure $P_{sm}=\rho^2 d\varepsilon(\rho, \delta=0)/d\rho$ of an isospin symmetric system wherein $\rho_n = \rho_p$. This added pressure depends strongly on the poorly constrained [34] density dependent term $S(\rho)$. The symmetry energy influences many properties of neutron stars. Besides contributing significantly to the pressure that counters the gravitational attraction, the symmetry energy determines the proton fraction, the pressure and density of the crust-core transition and other possible phase transitions within neutron stars, and has a large impact on neutrino cooling rates by Urca and modified Urca processes [35]. Astrophysical observations do not yet provide strong constraints on the symmetry energy. Since attaining a microscopic understanding of the EoS of dense matter constitutes an important objective of nuclear science [36], there have been ongoing experimental efforts to constrain the symmetry energy at various densities [11-23]. In the following, we combine constraints from the GW170817 event with laboratory constraints [19-21] to improve our understanding of the symmetry energy.

We start with the extraction of the neutron matter EoS from the GW170817 event [5]. During the inspiral phase of a neutron-star merger, the gravitational field of each neutron star induces a tidal deformation in the other [9]. The influence of the EoS of neutron stars on the gravitational wave signal during inspiral is contained in the dimensionless constant called the tidal deformability or tidal polarizability, $\Lambda = \frac{2}{3}k_2 \left(\frac{c^2 R}{GM}\right)^5$, where $G$ is the gravitational constant, $M$ and $R$ are the mass and radius of a neutron star and $k_2$ is the dimensionless Love number [5, 9] which is also sensitive to the compactness parameter ($M/R$). As the knowledge of the mass-radius relation uniquely determines the neutron-star matter EoS [37-38], information about EoS can be obtained from $\Lambda$.

In the original GW analysis of late-stage inspiral, an upper limit of $\Lambda < 800$ was obtained assuming 1.4 solar-mass neutron stars and a low spin scenario [5]. These values of $\Lambda$ are updated in a recent analysis to $300^{+420}_{-230}$ [7]. Requiring both neutron stars to have the same EoS led to even more restrictive $\Lambda$ values of $190^{+390}_{-120}$ and $R$ values of $11.9^{+1.4}_{-1.4}$ km [8]. Here we adopt a complementary approach of using laboratory observables and a neutron star model to advance our understanding of the microscopic nature of the EoS.

To calculate neutron star properties, we adopt the approach described in Ref. [39] to solve the Tolman-Openheimer-Volkov (TOV) equation. At $\rho<0.5\rho_0$, matter is inhomogeneous and must be described by a crustal EoS [40-46]. This low density region has no impact on the tidal polarizability, but it does significantly increases the stellar radius for stars with $R>13$ km, as shown below. At higher densities of $\rho \approx 0.5\rho_0 - 3\rho_0$, the matter in the neutron star is assumed to be homogeneous, nucleonic and beta-equilibrated. The pressure within this Fermi liquid neutron star core, $P=\rho^2 d\varepsilon(\rho, \delta\approx 1)/d\rho$ supports the star [38]. We connect this region to nuclear physics measurements by incorporating 245 well studied microscopic Skyrme effective interactions and their corresponding homogenious nuclear density functionals that have been constrained by nuclear experiments [31,47, 48] in the neutron star EoS. At $\rho>3\rho_0$ the stable phase of matter is unknown and strange or quark matter regions may predominate, depending on the EoS of nuclear matter. We use a polytropic EoS that smoothly matches to the Skyrme EoS at $3\rho_0$ to extend the stellar EoS to the central density region [38, 39] and to allow up to 2.17 solar-mass neutron stars to be supported. For the low mass neutron stars that are considered here, this high density region does not significantly influence our calculated neutron star properties. Without the polytropes, some of the Skryme interactions are not sufficiently repulsive at $\rho>3\rho_0$ to support a 2.17 solar-mass star. If we were to exclude them, the remaining Skyrmes would nevertheless demonstrate the general features of Figs 1 and 2.

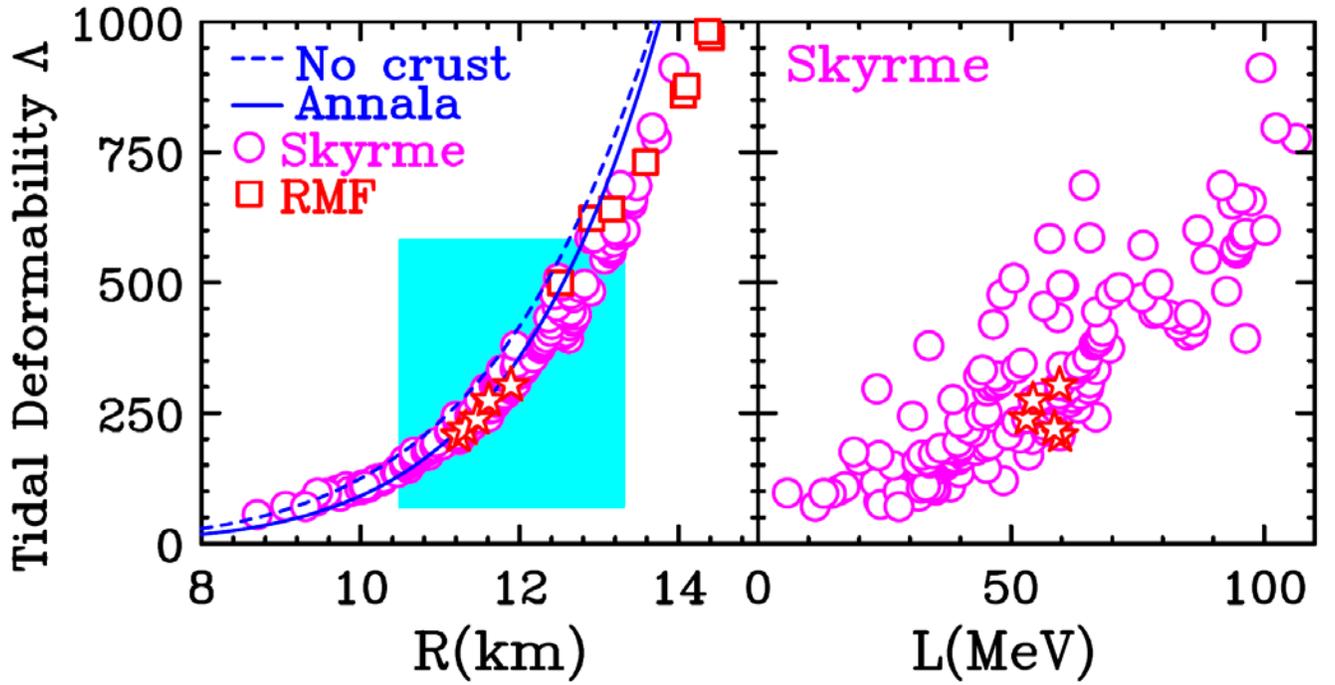

**Fig. 1:** (Left panel) Correlation between neutron-star tidal deformability and radii from current calculations (open circles) and from Ref. [10] (open squares). The light blue shaded area represents constraint from recent GW170817 analysis [8]. Five interactions, KDE0v1, LNS, NRAPR, SKRA, QMC700 deemed as the best in Ref. [31] in describing the properties of symmetric matter and calculated pure neutron matter are plotted as red stars. The solid curves is from Ref. [49] and the dashed curve is the best fit result if no crust is included in our neutron star model. (Right panel) Correlation between neutron-star tidal deformability and L.

For each of the Skyrme interactions that can support a 2.17 solar mass neutron star, we obtain a unique prediction for the neutron-star radius and tidal deformability that is represented by an open violet circle in Fig. 1. Our overall results are consistent with those represented by the open red squares constructed using relativistic mean-field interactions [10], but our radii at large $\Lambda$ exceed those of ref. [49]. The model in ref. [49] does not include a crust. If we neglect the crust in our calculations, we obtain the blue dashed curve, which resembles the blue solid curve from ref. [49]. Above $\Lambda > 600$, our calculations with a crust produce larger radii than do our calculations without a crust. This trend is consistent with ref. [50], which shows that the crust thickness increases inversely with neutron star compactness (M/R), but depends little on uncertainties in the crustal EoS.

The blue-shaded region, referred simply as "GW", indicates the allowed region of $\Lambda=70\text{-}580$ and $R = 10.5\text{-}13.3\ km$ obtained in [8]. Our calculations lie nearly diagonally across the box with about 130 interactions inside. Eleven conditions that describe the properties of symmetric matter and pure neutron matter were used in [31] to evaluate their Skyrme interactions. On average, the Skyrme interactions inside the GW box satisfy more than 8 of these constraints while those outside the box satisfy less than 6. We highlight the five interactions plotted in red stars: KDE0v1, LNS, NRAPR, SKRA, QMC700, which satisfy nearly all the 11 constraints. Their $\Lambda$ values (~250) with the associated radii (~11.3 km) are well within the GW constraint.

The interactions outside GW constraint tend to have very large (> 70 MeV) or very small L (<40 MeV) values. $L = 3\rho_0 dS(\rho)/d\rho\big|_{\rho_0}$ is proportional to the slope of the symmetry energy at saturation density and defines how rapidly the symmetry energy increases with density at $\rho_0$. Measurments of neutron skins [12] or asymmetry skins [47,51] provide constraints on L but the uncertainties in connecting L to $\Lambda$ can be large. This is shown by the large dispersion of the L vs. $\Lambda$ correlation in the right panel of Figure 1. This observation is also supported by calculations in Ref. [52].

To illustrate where the EoS must be known well to describe the GW constraint, we focus on two density regions, $0.67\rho_0$ and $2\rho_0$ where experimental data exists. The energy of neutron matter is experimentally best constrained at $0.67\rho_0$ [53] where an accurate value (~25 MeV) of the symmetry energy has been derived from the analysis of nuclei masses [48, 54] and isobaric analog states [47]. While the symmetry energy at sub saturation density is important for the crust-core transition in neutron star, it does not constrain the neutron star deformability, $\Lambda$, as shown in the left panel of Fig. 2.

At supra-saturation density, pressure at *2ρ₀* has been identified to be sensitive to the neutron-star radius [8, 52, 55]. At this density our calculations (open circles in the right panel of Fig. 2) show a strong correlation between the total pressure of each Skyrme functional and its calculated tidal deformability. This strong correlation combined with Λ-R correlation in Fig.1 implies a strong correlation between the neutron star radius and the neutron star pressure at *2ρ₀*, demonstrated previously by ref. [55].

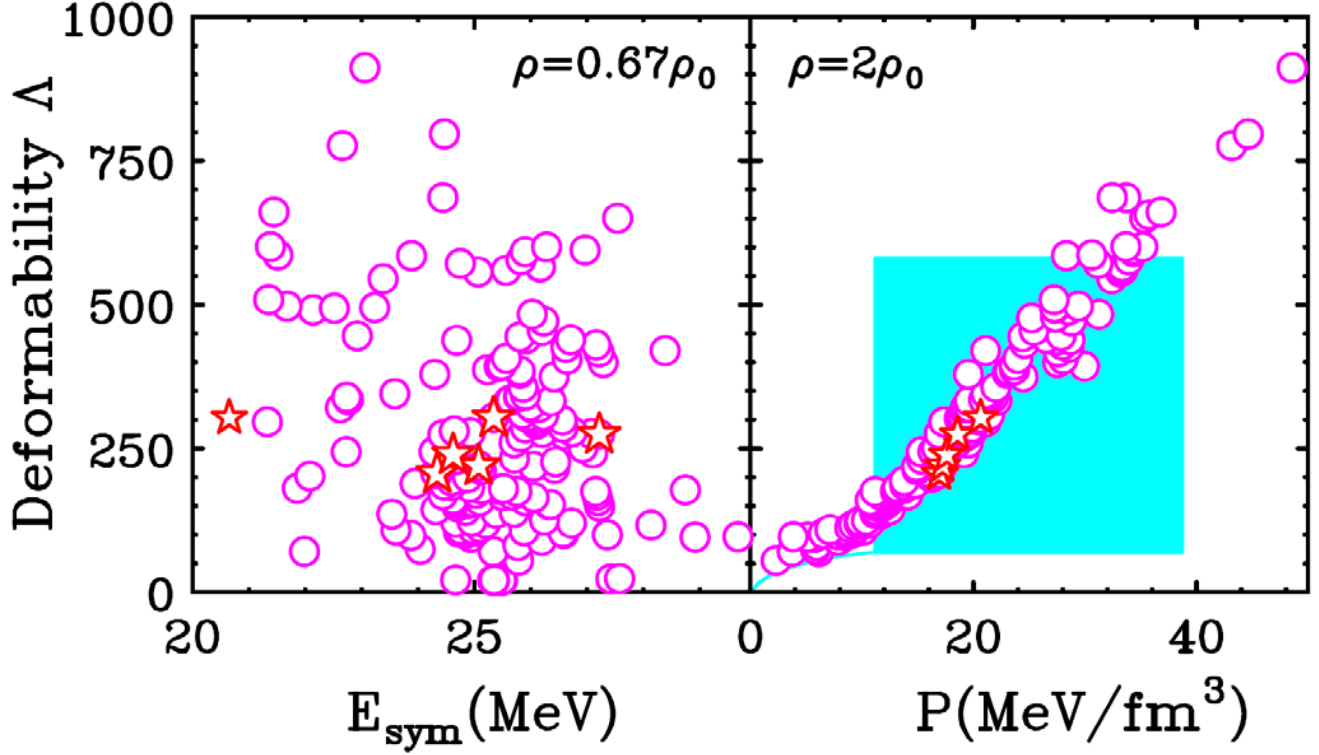

**Fig. 2:** Neutron-star tidal deformability vs. symmetry energy at *0.67ρ₀* (left panel) and deformability vs. pressure at *2ρ₀* (right panel).

Above the saturation density, the neutron star pressure has significant contributions from both the symmetry energy and the symmetric matter equations of state. The pressure constraints for symmetric matter have been obtained in Ref. [19] and confirmed in ref. [22] at densities ranging from *2ρ₀* to *4.5ρ₀* from the Heavy Ion (HI) measurements of collective flow in Au+Au collisions. Similar constraints on the pressure at densities ranging from *1.2ρ₀* to *2.2ρ₀* were obtained from kaon production measurements [20, 21]. The contours enclosed by blue solid and dotted lines in the left panel of Figure 3 represent these constraints on the symmetric matter EoS from flow [19] and kaon [20, 21] measurements, respectively. The contours in the HI constraints are at the 68% confidence level and include the uncertainties in the measurements as well as the theoretical uncertainties in extracting the EoS [19] from the measured data. Also shown are the 90% and 50% confidence level GW constraints on neutron matter published in [8] and represented by the light blue and green shaded areas respectively in Fig. 3.

Assuming the most probable values lie at the centers of these contours, we can deduce the dependence of the symmetry pressure, $P_{sym}=P_{nm}-P_{sm}=\rho^2 d(E_{sym}(\rho))/d(\rho)$ and how it increases with density as shown in the right panel of Fig. 3. For reference, we show two commonly used bracketing assumptions for the symmetry energy within neutron-stars suggested by Prakash et. al. [56]:

$$S(\rho)_{stiff}=12.7\ MeV\times(\rho/\rho_0)^{2/3}+38 MeV\times(\rho/\rho_0)^2/(1+\rho/\rho_0) \qquad (2)$$

$$S(\rho)_{soft}=12.7\ MeV\times(\rho/\rho_0)^{2/3}+19 MeV\times(\rho/\rho_0)^{1/2} \qquad (3)$$

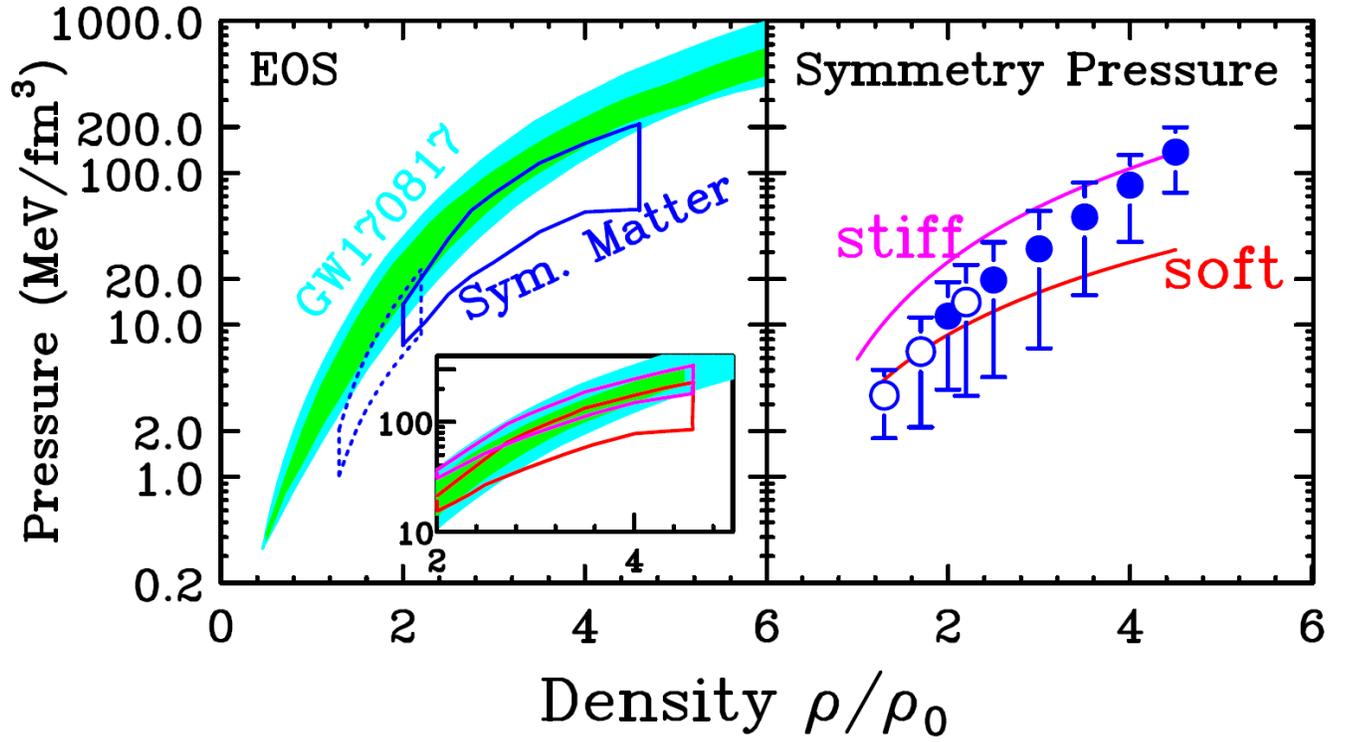

**Fig. 3:** (Left panel) Experimental and astrophysical constraints on equation of state in pressure vs. density. The shaded region represents the GW constraint [8], after converting the original unit for pressure of dyn/cm$^2$ to MeV/fm$^3$ and the for density to units of saturation density, $\rho_0=2.74\times10^{14}$ g/cm$^3$, to allow direct comparisons to nuclear physics constraints. Dashed contours display constraints for symmetric matte from flow measurements [19]. (Right panel) Symmetry pressure as a function of density extracted from the GW [8] and flow [19] constraints. The upper (labelled as stiff) and lower (labelled as soft) lines correspond to the symmetry pressure (for $\delta=1$) from Eqs. (2) and (3), respectively.

For convenience, we label Eq. (2) and Eq. (3) as "stiff" and "soft", respectively. The derived symmetry pressures are plotted as violet (stiff) and red (soft) solid curves in the right panel of Figure 3. The density dependence increases rapidly. At low density, the most probable symmetry pressure appears closer to the "soft" symmetry energy. At $4.5\rho_0$, the data seems to agree better with the "stiff" symmetry energy. However, the uncertainties are very large. The error bars are obtained from the combination, in quadrature, of the 50% confidence level boundaries for $P_{nm}$ (green shaded area) and $P_{sm}$ extracted from the HI and kaon contours [57]. For reference, the total pressure contours obtained by adding the stiff and soft symmetry pressures to the experimental symmetric matter pressure contours as described in ref. [19] are plotted as the corresponding violet (stiff) and red (soft) contours in the inset in the left panel. While the constraint on $P_{sym}$ is within the expected bounds, it is clear that more accurate measurements designed to isolate the symmetry energy or symmetry pressure are needed.

In summary, we have calculated neutron star properties using well characterized Skyrme interactions in order to learn how to constrain the symmetry energy at supra-saturation densities. At sub-saturation density, the dipole deformability is not sensitive to experimental observable such as nuclei masses. The recent GW constraint excludes the Skyrme interactions with extreme values of L, the slope of the symmetry energy related to the neutron matter pressure at saturation density. The calculations also suggest that the neutron star properties can be most sensitively probed around twice the saturation density. By combining the GW and existing constraints on the symmetric matter EoS from heavy ion measurements, we obtain the most probable value for the density dependence of the symmetry pressure between 1.2 to 4.5 times the saturation density. As the precision of GW constraints on neutron matter and both constraints on symmetry energy and heavy ion constraints on symmetric matter improve, more stringent constraints on the symmetry energy will be obtained.

**Acknowledgement**:

We would like to thank Prof. C. J. Horowitz for advice and fruitful discussions. This work was partly supported by the US National Science Foundation under Grant PHY-1565546 and by the U.S. Department of Energy (Office of Science) under Grants DE-SC0014530, DE-NA0002923, DE-FG02-87ER40365 (Indiana University) and DE-SC0018083 (NUCLEI SciDAC-4 Collaboration). This work was stimulated by discussions with participants at the INT-JINA Symposium "First multi-messenger observations of a neutron-star merger and its implications for nuclear physics"